\documentclass[aps,prl,twocolumn,superscriptaddress,nofootinbib]{revtex4-1}
\usepackage[latin1]{inputenc}
\usepackage{graphicx}
\usepackage{xcolor}
\usepackage{amsmath}
\usepackage{ulem}
\begin{document}
\title{Emergent Strain-Stiffening in Interlocked Granular Chains}
\author{Denis Dumont}
\affiliation{Laboratoire Interfaces $\&$ Fluides Complexes, Universit\'e de Mons, 20 Place du Parc, B-7000 Mons, Belgium.}
\author{Maurine Houze}
\affiliation{Laboratoire Interfaces $\&$ Fluides Complexes, Universit\'e de Mons, 20 Place du Parc, B-7000 Mons, Belgium.}
\author{Paul Rambach}
\affiliation{Laboratoire Interfaces $\&$ Fluides Complexes, Universit\'e de Mons, 20 Place du Parc, B-7000 Mons, Belgium.}
\affiliation{Laboratoire de Physico-Chimie Th\'{e}orique, UMR CNRS Gulliver 7083, ESPCI Paris, PSL Research University, 10 rue Vauquelin, 75005 Paris, France.}
\author{Thomas Salez}
\affiliation{Laboratoire de Physico-Chimie Th\'{e}orique, UMR CNRS Gulliver 7083, ESPCI Paris, PSL Research University, 10 rue Vauquelin, 75005 Paris, France.}
\affiliation{Univ. Bordeaux, CNRS, LOMA, UMR 5798, F-33405 Talence, France.}
\affiliation{Global Station for Soft Matter, Global Institution for Collaborative Research and Education, Hokkaido University, Sapporo, Japan.}
\author{Sylvain Patinet}
\affiliation{PMMH, ESPCI Paris/CNRS-UMR 7636/Univ. Paris 6 UPMC/Univ. Paris 7 Diderot,
PSL Research Univ., 10 rue Vauquelin, 75231 Paris cedex 05, France.}
\author{Pascal Damman}
\email{pascal.damman@umons.ac.be}
\affiliation{Laboratoire Interfaces $\&$ Fluides Complexes, Universit\'e de Mons, 20 Place du Parc, B-7000 Mons, Belgium.}
\date{\today}
\begin{abstract}
Granular chain packings exhibit a striking emergent strain-stiffening behavior despite the individual looseness of the constitutive chains. Using indentation experiments on such assemblies, we measure an exponential increase in the collective resistance force $F$ with the indentation depth $z$, and with the square root of the number $\mathcal{N}$ of beads per chain. These two observations are respectively reminiscent of the self-amplification of friction in a capstan or in interleaved books, as well as the physics of polymers. The experimental data are well captured by a novel model based on these two ingredients. Specifically, the resistance force is found to vary according to the universal relation: $\log F \sim \mu \sqrt{\mathcal{N}} \Phi^{11/8}z/ b $, where $\mu$ is the friction coefficient between two elementary beads, $b$ is their size, and $\Phi$ is the volume fraction of chain beads when semi-diluted in a surrounding medium of unconnected beads. Our study suggests that theories normally confined to the realm of polymer physics at a molecular level can be used to explain phenomena at a macroscopic level. 
This class of systems enables the study of friction in complex assemblies, with practical implications for the design of new materials, the textile industry, and biology.
\end{abstract}
\maketitle

In nature and architecture, a wide variety of stable structures are formed from dense assemblies of randomly-distributed objects~\cite{insect,archi}, as beautifully illustrated by the various shapes of bird nests~\cite{nest}. The aggregates can be made of elementary components with arbitrarily complex and sometimes living shapes, as in the case of fire ants~\cite{ant}. Most physical studies focus on simpler but highly anisotropic objects, {\it e.g.} rods, that ensure a solid-like collective behavior and thus the great stability of their assembly~\cite{philipse_1996,trepanier_2010}, even under large compressive stresses. However, there is no topological constraint between the constitutive objects that could hinder their motion within the aggregate. The mechanical stability of rod aggregates is thus intimately related only to the packing~\cite{nagel} and the solid friction~\cite{friction_book} at the contact points. Further works have been devoted to the mechanics of assemblies made of more complex objects, such as stars~\cite{star}, or U- and Z-shaped particles~\cite{U_Z_particle1,U_Z_particle2}. In those cases, in contrast to rods, it is considered that topological constraints are the key ingredient for the observed increase in rigidity. 

Apart from using rods or rigid non-convex objects, a third approach to produce cohesive assemblies from non-interacting elements is to aggregate soft slender objects, such as fibers~\cite{galilee,bayman,keller} or granular chains~\cite{granular_chain}. In the latter case, a strain-stiffening phenomenon was reported experimentally in the absence of dilution, and was qualitatively related to the formation of loops between chains~\cite{brown}. However, a quantitative theoretical mechanism was still lacking, despite the fact that the jamming of these systems might have some fundamental connexion with the glass transition of polymeric materials. Besides, while the packing of rigid rods is drastically affected by their aspect ratio, granular chains exhibit less impressive changes. For very thin rods, packing fractions as low as 0.2 have been reported and theoretically described through the random-contact equation~\cite{philipse_1996}, while for granular chains the lowest packing fractions range between 0.4 and 0.5~\cite{granular_chain}. This reduced influence of the shape anisotropy might be related to the high flexibility of granular chains.

In this Letter, we investigate the emergent mechanical rigidity of granular chain assemblies. Performing indentation experiments, we observe an exponential growth of the resistance force with the product of the indentation depth by the square root of the number of beads per chain. The first factor is reminiscent of the self-amplification of friction in a capstan~\cite{capstan}, as well as in interleaved books~\cite{alarcon_2016,Dalnoki_2016}, where the increase in resistance is created by, and proportional to, the force exerted by the operator. The second factor points towards the central role played by inter-chain topological constraints, as in polymer physics~\cite{degennes}. Therefore, we propose a new interlocking model based on these two ingredients, and confront it to the experimental data. We find a very good agreement, including for the extended situation of semi-dilute chains within a surrounding medium of unconnected beads. 

\begin{figure}
\begin{center}
\includegraphics[width=1\linewidth]{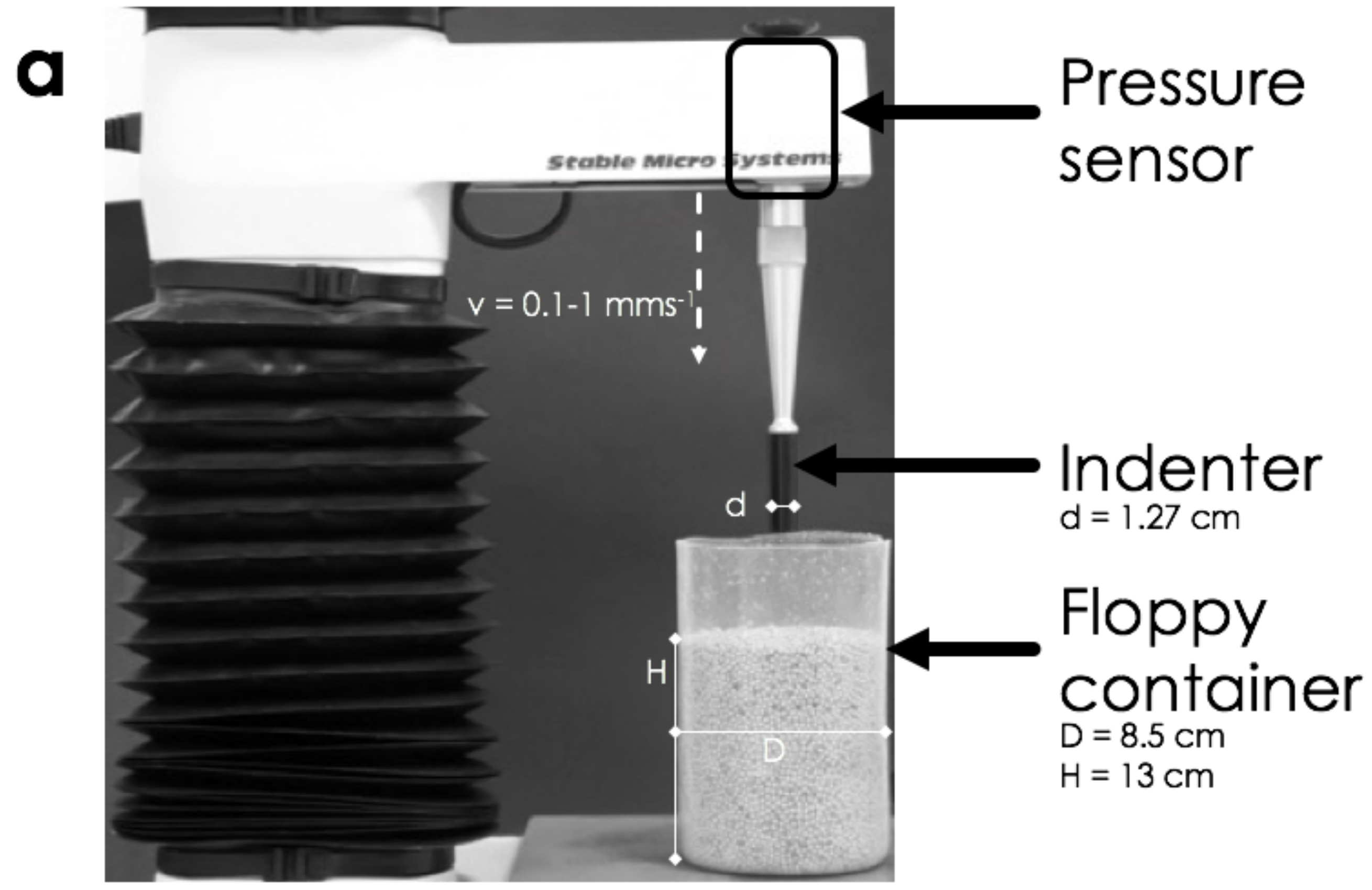}
\includegraphics[width=1\linewidth]{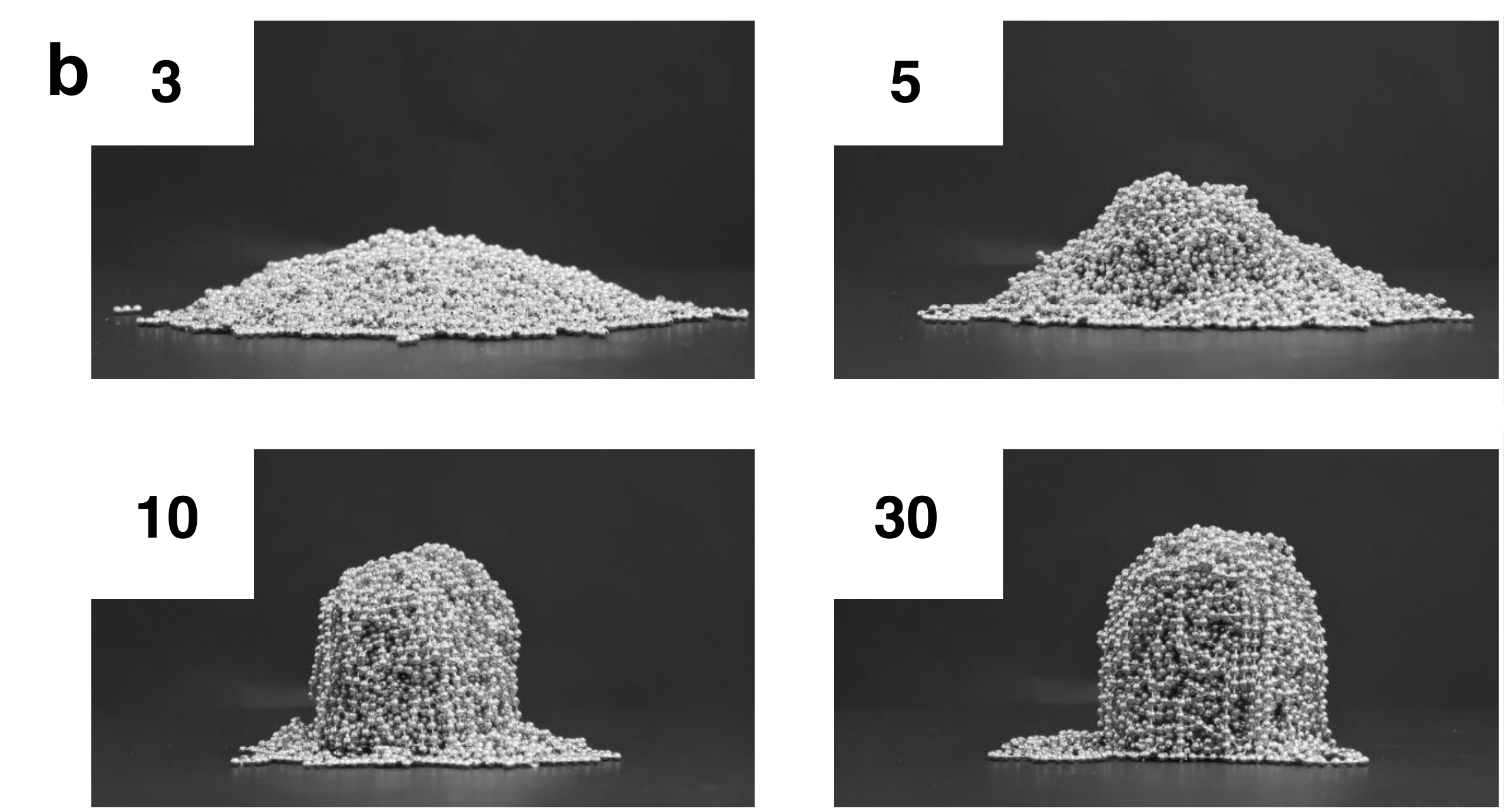}
\end{center}
\caption{(a) Description of the experimental setup (see main text for details). (b) Snapshots showing the final states of dense ($\Phi=1$) granular chain assemblies after the removal of their cylindrical container, for various numbers $\mathcal{N}$ of beads per chain as indicated in the top left corner of each snapshot.}
\label{Fig1}
\end{figure}

Our experiments are conducted on large, monodisperse, assemblies of granular chains, each chain being made of $\mathcal{N}$ connected steel beads of diameter $b=2\,\text{mm}$, with $2\le \mathcal{N} \le 50$. The random packing of the ensemble is achieved through sharp taps until a constant packing density is reached, in agreement with~\cite{granular_chain}. Using a ProbeTack system (Fig.~\ref{Fig1}(a)), the force needed to indent the assembly in its cylindrical container is measured as a function of the indentation depth $z$. We use a cylindrical indentor with diameter $d=1.27\, \text{cm}$, that is displaced at constant velocity. Low velocities, in the $0.1-1\,\text{mm}\, \textrm{s}^{-1}$ range, are used in order to ensure a quasistatic regime and to avoid velocity-dependent drag forces~\cite{durian}. We employ floppy containers, molded with soft crosslinked Sylgard, in order to limit vaulting and arching, and thus prevent the Janssen effect~\cite{andreotti}. Semi-dilute systems are also studied, and prepared by mixing monodisperse granular chains ($\mathcal{N}=30$) with unconnected beads, resulting in a volume fraction $\Phi=V_{\textrm{chains}}/(V_{\textrm{chains}}+V_{\textrm{unconnected}})$ of chain beads, where $V_{\textrm{chains}}$ and $V_{\textrm{unconnected}}$ are the total volumes of chain beads and unconnected beads respectively, before mixing (we neglect the volume of mixing).

We first start with $\Phi=1$, \textit{i.e.} with pure granular chain assemblies. As qualitatively shown in Fig.~\ref{Fig1}(b), these assemblies exhibit a behavior similar to rod aggregates~\cite{trepanier_2010}: when initially packed in a cylindrical container, short-chain assemblies collapse after removal of the container and form a conical pile, like bare sand, while long-chain assemblies are surprisingly able to retain their cylindrical shape despite the looseness of each individual chain.
 
\begin{figure}
\begin{center}
\includegraphics[width=1\linewidth]{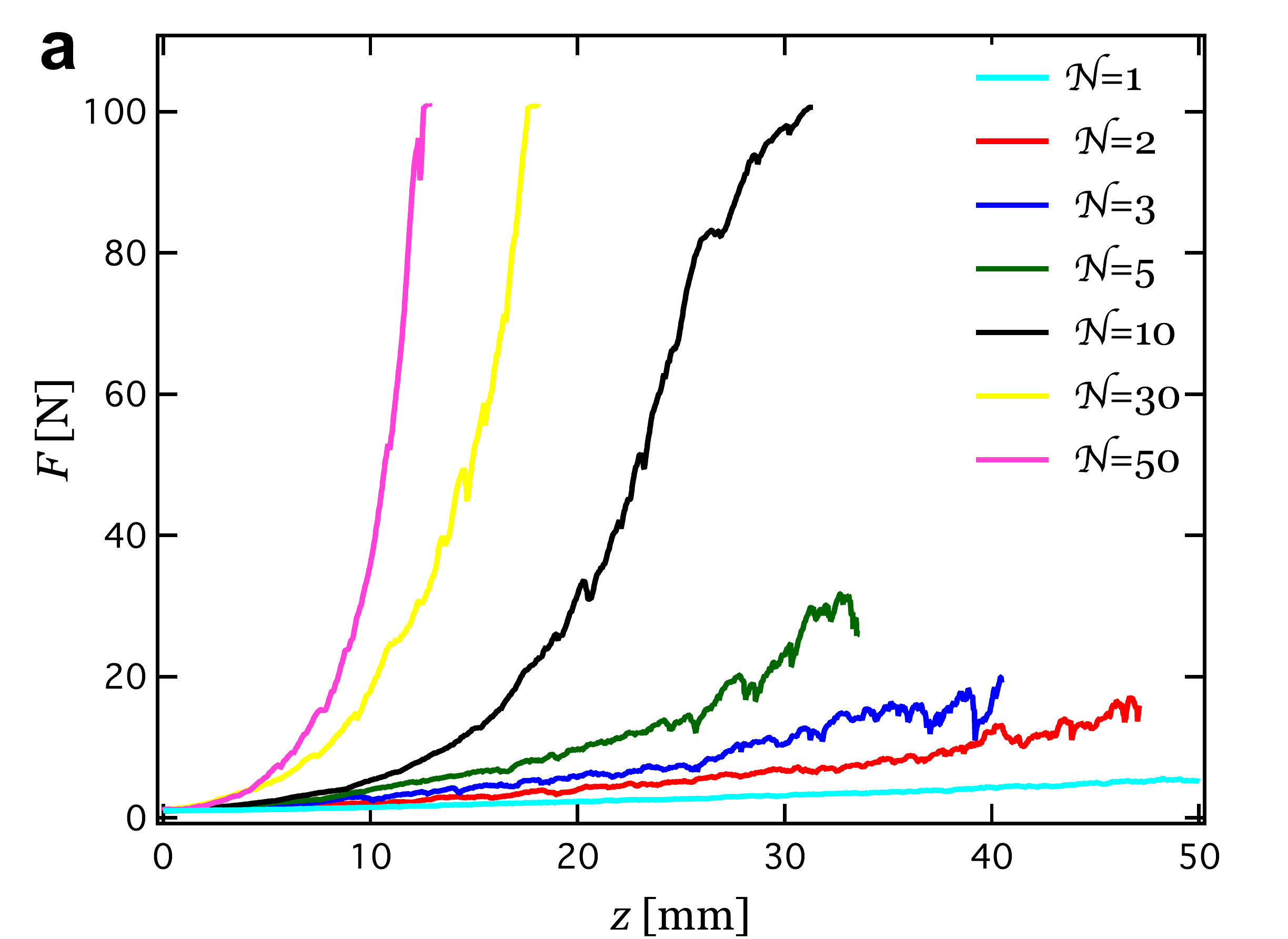}
\includegraphics[width=\linewidth]{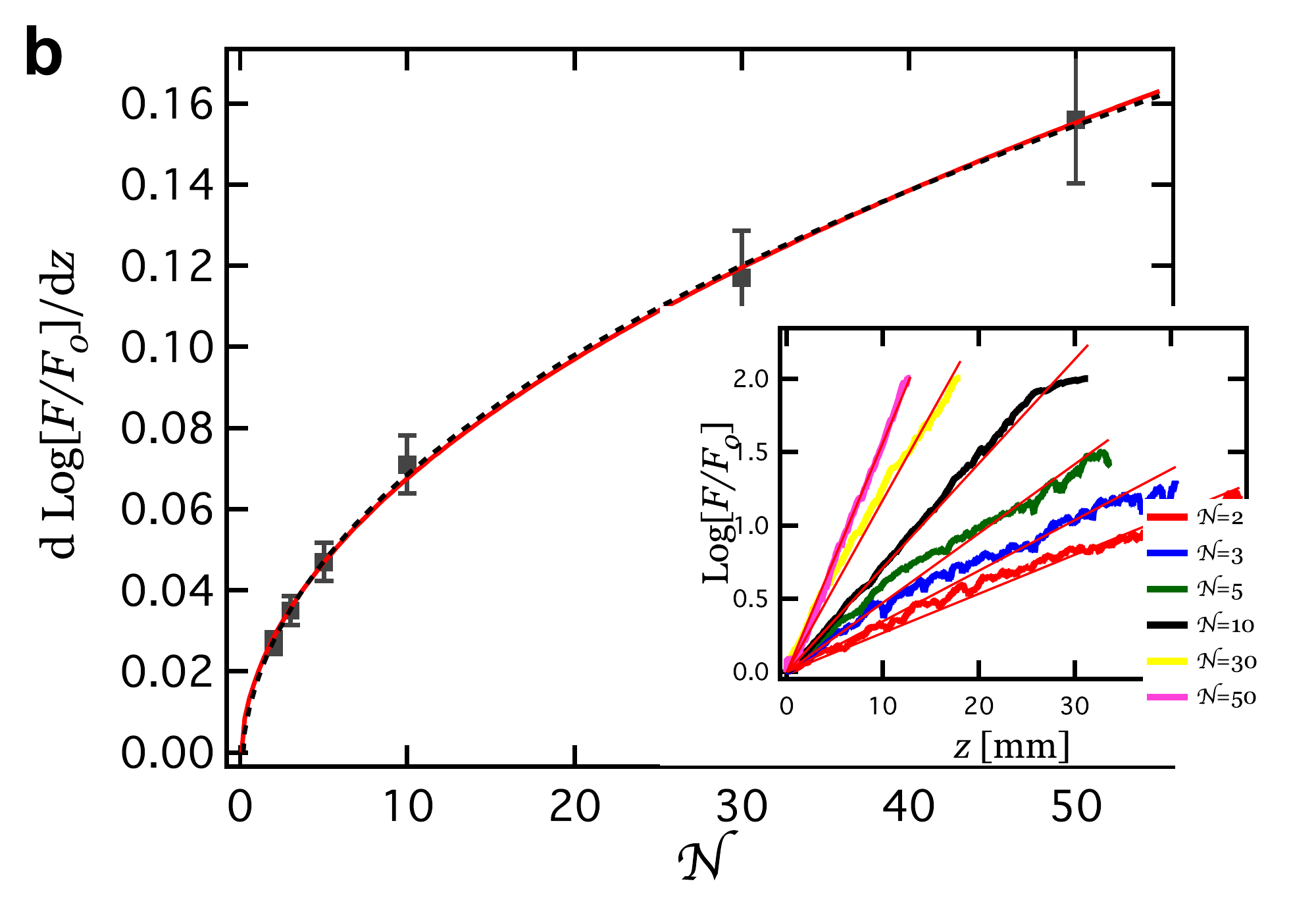}
\end{center}
\caption{(a) Indentation force $F$ as a function of indentation depth $z$, for granular chain assemblies with various numbers $\mathcal{N}$ of beads per chain as indicated. (b) Slope of the log-linear representation of the $\mathcal{N}\geq2$ data (see current inset) as a function of  $\mathcal{N}$. The solid line indicates a $\sim \sqrt{N}$ behaviour, while the dashed one corresponds to the best power-law fit: $0.03 \, \mathcal{N}^{0.46\pm 0.06}$. (Inset) Log-linear representation of the $\mathcal{N}\geq2$ data, with $F_0 = 1$~N an arbitrary reference force. The solid lines indicate linear trends in this representation.}
\label{Fig2}
\end{figure}

To understand this striking emergent rigidity, we systematically study the influence of the depth $z$ and the number $\mathcal{N}$ of beads per chain, on the force $F$ needed to indent such an assembly. The resulting motion of the indentor involves local plastic-like structural rearrangements~\cite{andreotti}, characterized by small force discontinuities, that we do not discuss further here. Figure~\ref{Fig2}(a) summarizes our raw results. The indentation force continuously increases with the depth. For unconnected beads ($\mathcal{N}=1$) and small indentation depths, a nearly-linear force-displacement behavior is observed, in agreement with~\cite{Peng_2009}. A sudden increase of the force has been however reported when the indentor gets very close to the bottom of the container~\cite{expo_bead}. 
For $\mathcal{N}\geq2$, the force increase is not compatible anymore with an effective Hookean behavior. Indeed, the stiffness, given by the slope $dF/dz$, increases with $z$ thus confirming the strain-stiffening scenario previously reported~\cite{brown}. In addition, the force and stiffness both increase sharply with the number $\mathcal{N}$ of beads per chain, indicating the predominant influence of chain connectivity. 
As shown in Fig.~\ref{Fig2}(b), the experimental curves are consistent with an exponential growth of the force with the indentation depth. Furthermore, the dependence in the number of beads per chain appears through the slope $\textrm{d\,log}(F)/\textrm{d}z$ given in Fig.~\ref{Fig2}(b) which can be fitted by the power law $\sim\mathcal{N}^{0.46}$. To summarize, the observed mechanical stiffness of the granular chain assembly can thus essentially be described by the relation $\log F \propto z \sqrt{\mathcal{N}}$. Hereafter, we propose an interlocking model based on two main ingredients that allows to rationalize these observations.

First, since we operate in a quasistatic regime there is no drag force, and the resistance to indentation should be only related to internal friction forces hindering the motion of the constitutive objects~\cite{durian}. Previous studies about self-amplified friction in complex assemblies have shown that the frictional resistance essentially grows exponentially with the relevant control parameter, \textit{e.g.} the angle in a capstan~\cite{capstan}, or the Hercules number~\cite{alarcon_2016,Dalnoki_2016} in interleaved books. This general behaviour is intimately rooted in the fact that the increase in the force exerted by the operator is proportional to that force itself and to an effective coefficient of friction induced by the specific geometry. Indeed, the applied force generates normal loading on a certain amount of locking points in the system, which in turn induces a resistive friction force according to Amontons-Coulomb law at the onset of motion. For our granular chains, the motion of the indentor becomes possible as soon as the locking points on one chain are released, and we thus postulate a geometrically-induced self-amplification of friction described by the equation:
\begin{equation}
\frac{\textrm{d}F}{\textrm{d}z} \sim \mu \ F {M_z}\ ,
\label{eq1}
\end{equation}
where $\mu$ is the friction coefficient between two elementary beads, and where $M_z$ is the number of locking points on one chain per unit length of indentation.

Secondly, inspired by Edwards' general conjecture for athermal granular matter~\cite{edwards}, we assume that randomly-packed granular chains behave as thermally equilibrated flexible polymers. In addition, we focus only on the large-$\mathcal{N}$ asymptotics.
In this framework, the chains follow the ideal random-walk statistics~\cite{degennes} with a typical radius of gyration $R_0\sim\sqrt{\mathcal{N}} \,b$. We stress that self-avoiding effects indeed cancel in polymer melts ($\Phi=1$) due to the screening of excluded-volume interactions. Since the chain assembly is dense, the volume $v\sim\mathcal{N}^{3/2} b^3$ pervaded by a given chain is in fact occupied by $v/(\mathcal{N} b^3)\sim\sqrt{\mathcal{N}}$ chains, on average. As a corollary, non-bonding contacts between beads of the assembly are rarely (probability $\sim 1/\sqrt{\mathcal{N}}$) intrachain contacts and thus mostly interchain contacts. Since all, and only, the non-bonding contacts are frictional locking points, quantifying $M_z$ simply amounts to estimating the number of interchain contacts -- the so-called interlocking points from now on -- per chain and per unit length of indentation. As each of the $\mathcal{N}$ beads of a given chain corresponds to $\sim 1$ interlocking point, one has:
\begin{equation}
M_z\sim \frac{\mathcal{N}}{R_0}\sim \frac{\sqrt{\mathcal{N}}}{b}\ .
\label{eq2}
\end{equation}

Then, inserting Eq.~(\ref{eq2}) in the solution of Eq.~(\ref{eq1}) leads to:
\begin{equation}
\log\left(\frac{F}{F_0}\right)\sim \frac{\mu \sqrt{\mathcal{N}}z}{b}\ ,
\label{eq3}
\end{equation}
which is the observed form in Fig.~\ref{Fig2}(b), with $F_0 = 1$~N an arbitrary reference force. Since we remained at the level of scaling, we missed a prefactor in the right-hand side of Eq.~(\ref{eq3}). However, this prefactor can be estimated from the fit of Fig.~\ref{Fig2}(b), knowing $b=2\,\textrm{mm}$ and $\mu$. As no specific cleaning procedure was applied to our chains, a thin layer of lubricant could likely remain between beads, and $\mu$ should thus be close to $0.1$, bringing the missing prefactor around $0.6$. We have thus demonstrated that our interlocking model, based on both self-amplified friction and polymer chain statistics, is compatible with the $\mathcal{N}\geq2$ data in the pure case ($\Phi=1$). 

To avoid any coincidence, and validate the model further, we finally test the polymer analogy by considering the dilution of granular chains ($\mathcal{N}=30$) with unconnected beads. Depending on the volume fraction $\Phi$, polymer solutions can be in 
either~\cite{degennes}: {\it (i)} the dilute regime ({\it i.e.} very low volume fractions), where the chains are isolated from each other. There, the relevant length scale is the size of a swelled chain, given by Flory's radius $R_{\textrm{F}}\sim \mathcal{N}^{3/5}b$. We stress that the $3/5$ exponent reflects the importance of self-avoiding effects in dilute solutions, in contrast to the previous melt-like case; or {\it (ii)} the semi-dilute regime ({\it i.e.} high volume fractions), where the chains start to interpenetrate. This regime is characterized by a new length scale $\xi(\Phi)$, the blob size, with $b\le \xi \le R_{\textrm{F}}$. The transition between the two regimes occurs at a critical volume fraction $\Phi^*\sim \mathcal{N}b^3/R_{\textrm{F}}^{\,3}\sim\mathcal{N}^{-4/5}$. We specifically prepare our system in the semi-dilute regime, in order to maintain some interchain contacts, and the associated self-amplification of friction. For $\mathcal{N}=30$ granular chains, this requires that $\Phi> \Phi^*\approx 0.06$.

Essentially, the self-amplified friction mechanism is maintained while the dilution of chains reduces the number of interlocking points. Thus, Eq.~\ref{eq1} remains valid provided the number of interlocking points on one chain per unit length $M_z$ is adjusted. It must range from the expression in Eq.~\ref{eq2} at the melt-like volume fraction $\Phi\sim1$, down to $\sim1/R_{\textrm{F}}$ at the critical volume fraction $\Phi\sim\Phi^*$. 
\begin{figure}
\includegraphics[width=0.75\linewidth]{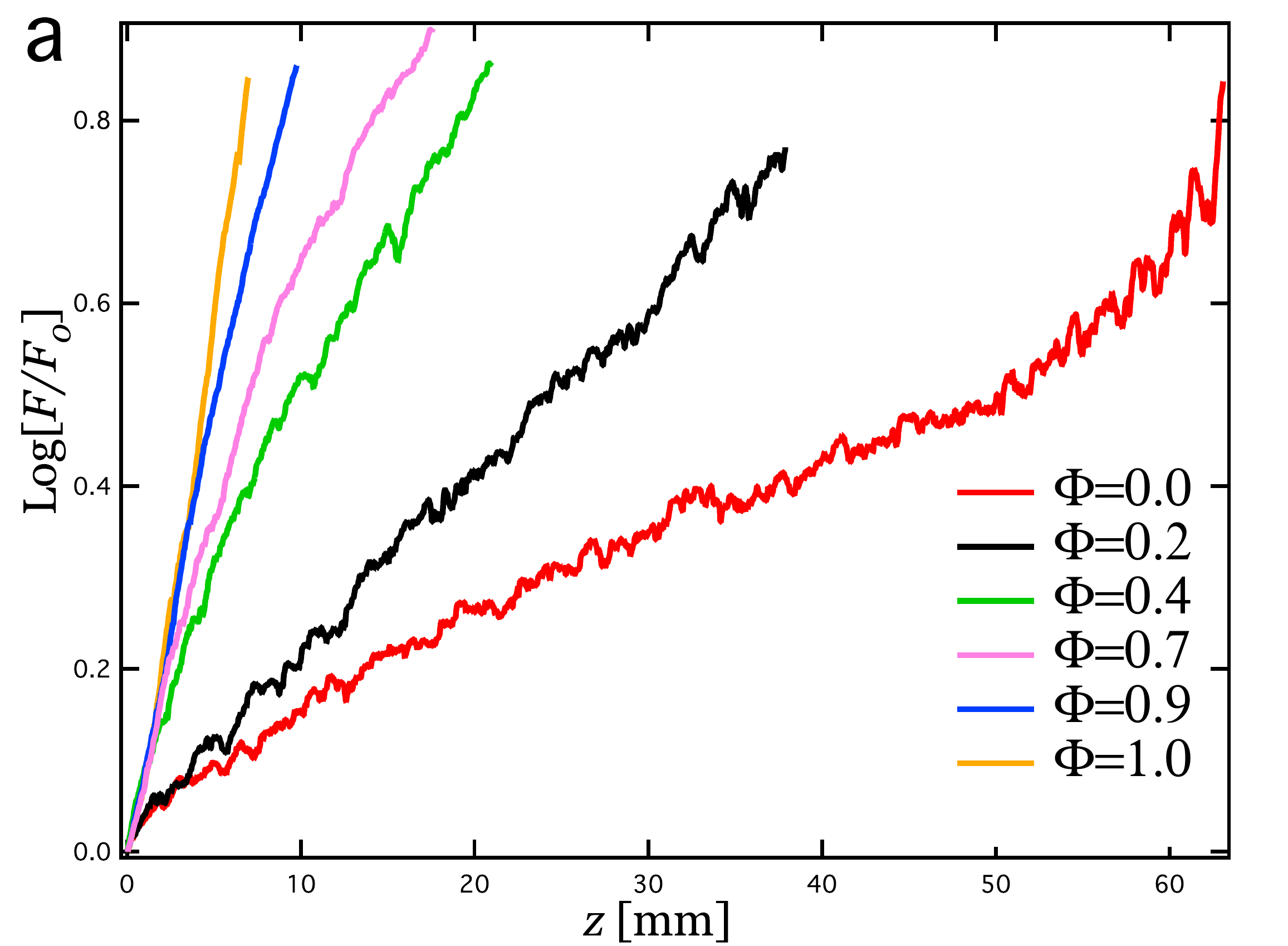}
\includegraphics[width=0.75\linewidth]{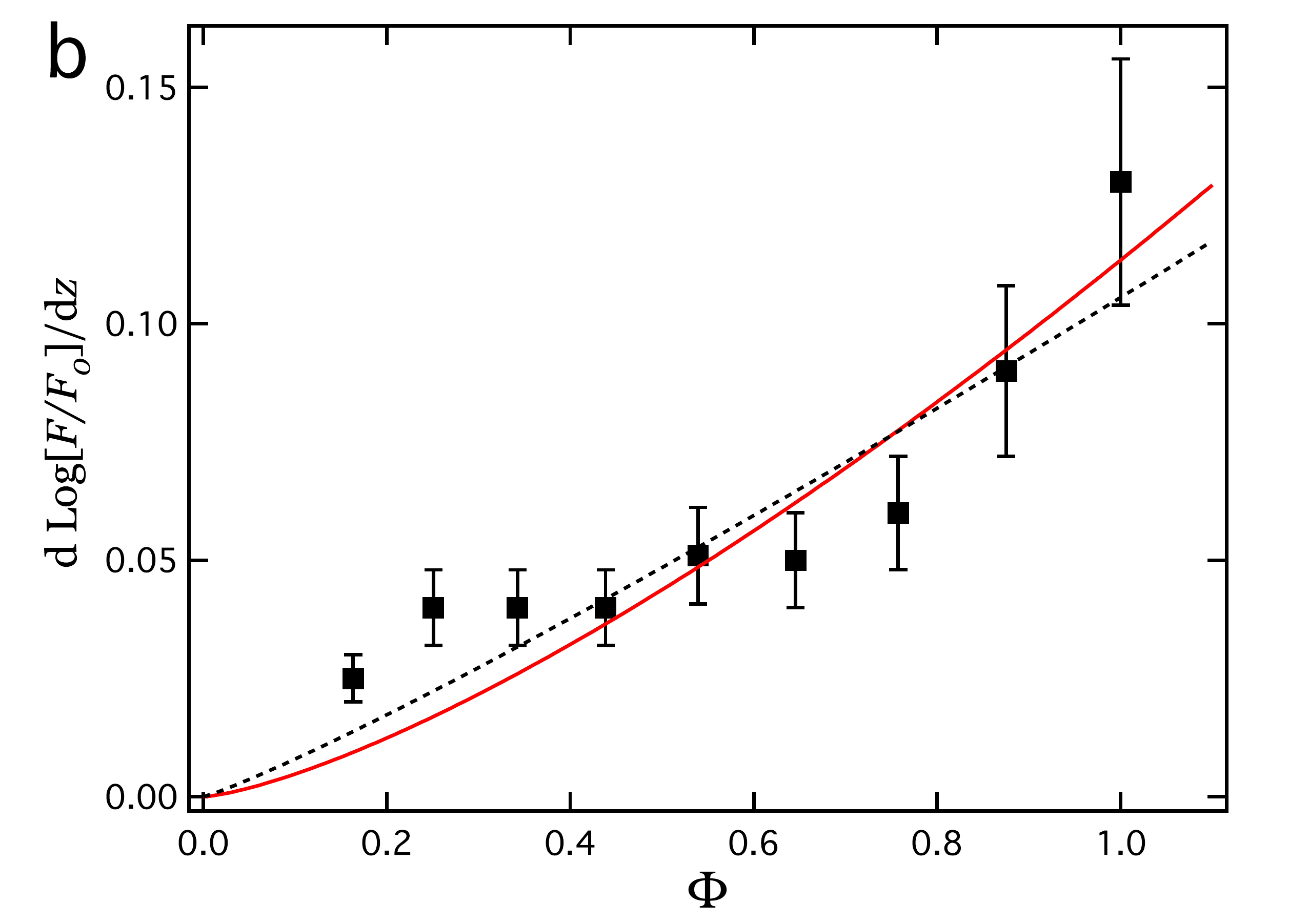}
\caption{(a) Log-linear representation of the indentation force $F$ as a function of the indentation depth $z$, for granular chain assemblies with $\mathcal{N}=30$ beads per chain, semi-diluted in a medium of unconnected beads, with selected (for clarity) volume fractions $\Phi$ of chain beads as indicated. (b) Slope calculated in the log-linear representation (a) as a function of the volume fraction $\Phi$ of chain beads. The solid line indicates a $\sim \Phi^{11/8}$ behaviour, while the dashed one corresponds to the best power-law fit: $\sim\Phi^{1.1 \pm 0.2}$.}
\label{Fig3}
\end{figure}
According to the physics of semi-dilute polymer solutions in a good -- athermal -- solvant~\cite{degennes}, each chain can be viewed as a succession of $\mathcal{N}/g$ independent blobs containing a subset of $g$ connected beads each. In such a renormalized picture, the whole solution can be viewed as a dense melt of ideal blob chains, which implies $\Phi\sim gb^3/\xi^3$. The portion of a chain inside a blob is in a purely dilute state (\textit{i.e.} with only unconnected beads around) and thus has a self-avoiding conformation leading to $\xi\sim g^{3/5}b$. As each of the $\mathcal{N}/g$ blobs of a given ideal chain of blobs corresponds to $\sim 1$ interlocking point, Eq.~\ref{eq2} must be replaced by:
\begin{equation}
M_z \sim {\sqrt{\mathcal{N}/g} \over \xi} \sim {\sqrt{\mathcal{N}} \Phi^{11/8} \over b}  \ ,
\label{eq4}
\end{equation}
that has the correct limiting values mentioned above. 

Then, inserting Eq.~(\ref{eq4}) in the solution of Eq.~(\ref{eq1}) leads to the universal relation:
\begin{equation}
\log \left({F \over F_0}\right) \sim \frac{\mu \sqrt{\mathcal{N}} \Phi^{11/8}z}{ b}\ ,
\label{eq5}
\end{equation}
that is valid for all $\Phi$ such that $\Phi^*<\Phi<1$, and for which Eq.~(\ref{eq3}) is a limiting case when $\Phi\rightarrow1$. The reference force $F_0$ and the missing prefactor have already been discussed after Eq.~(\ref{eq3}) and are not modified here. 

\begin{figure}
\begin{center}
\includegraphics[width=\linewidth]{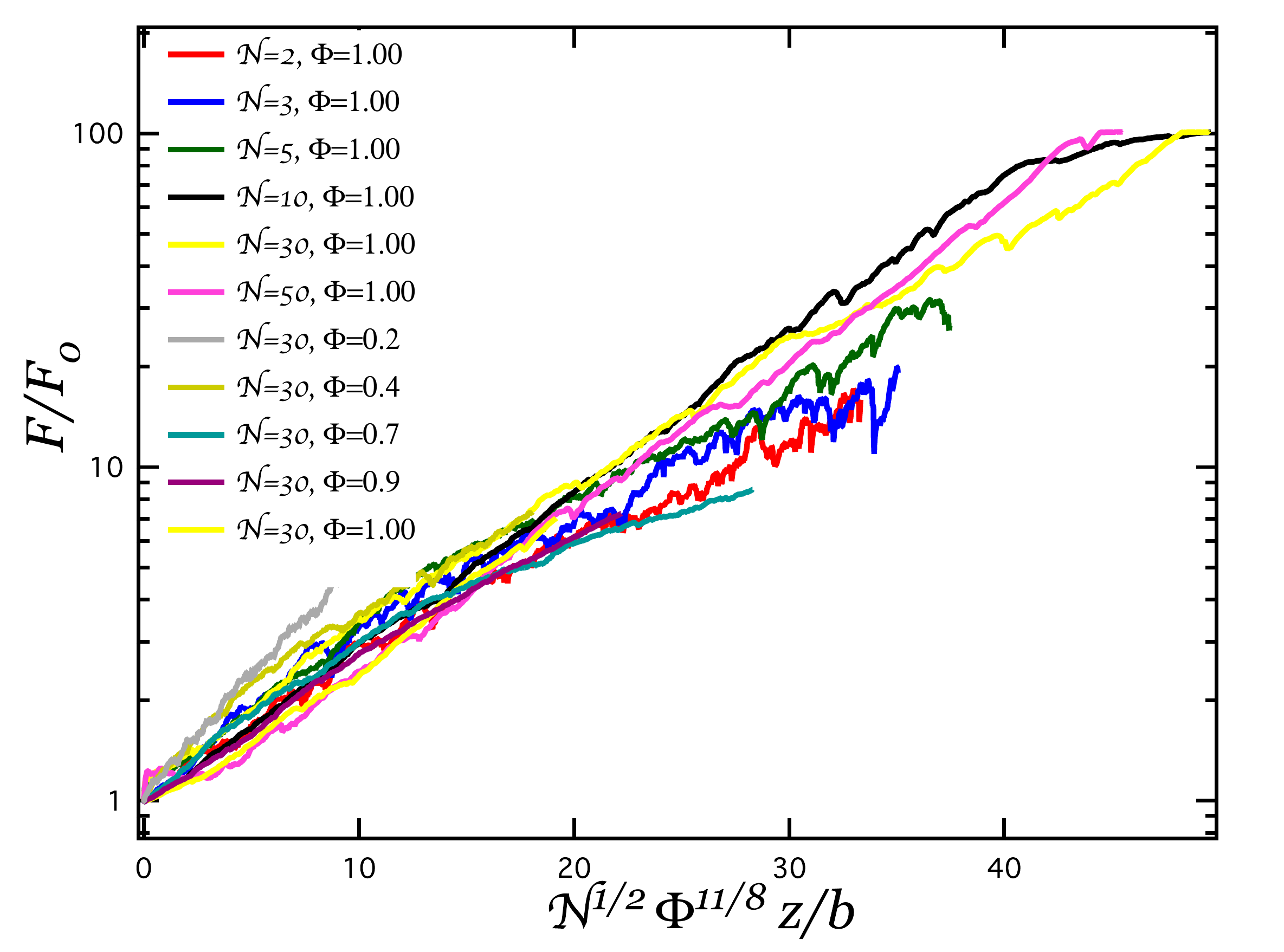}
\end{center}
\caption{Universal log-linear representation of all the dense and semi-dilute granular chain indentation experimental data of Figs.~\ref{Fig2}(inset) and~\ref{Fig3}(a), corresponding to $\mathcal{N}\geq2$ and $\Phi>\Phi^*\sim0.06$, as predicted by Eq.~(\ref{eq5}). The ensemble corresponds to various numbers $\mathcal{N}$ of beads per chain and volume fractions $\Phi$ of chain beads as indicated. Note that the dimensionless friction coefficient $\mu$ is not included in the $x$-axis, as it is neither varied nor measured, and since there is anyway a missing prefactor in Eq.~(\ref{eq5}).}
\label{Fig4}
\end{figure}

As shown in Fig.~\ref{Fig3}(a), the measured indentation force $F$ in the semi-dilute regime strongly increases with the volume fraction $\Phi$ of chain beads, while keeping the previous exponential behavior with indentation depth $z$. Similarly to Fig.~\ref{Fig2}(b), we plot the corresponding slope $\textrm{d\,log}(F)/\textrm{d}z$ as a function of $\Phi$ in Fig.~\ref{Fig3}(b), and find a best power-law fit of $\textrm{d\,log}(F)/\textrm{d}z\sim\Phi^{1.1}$, with an exponent close to the theoretical value $11/8$.
Following the prediction of Eq.~(\ref{eq5}), all the experimental data of Figs.~\ref{Fig2}(b)(inset) and~\ref{Fig3}(a) should belong to a single master curve. The universal collapse observed in Fig.~\ref{Fig4} corroborates the proposed interlocking mechanism.

In conclusion, the emergent strain-stiffening behavior in granular chain assemblies, as observed through indentation experiments, seems to originate from a self-amplification of friction due to polymer-like interlocking contacts. Interestingly, this system exhibits a new self-amplification exponent ($1/2$, in chain length) with respect to previous exponents for capstan ($1$, in angle)~\cite{capstan} and interleaved books ($2$, in number of sheets)~\cite{alarcon_2016,Dalnoki_2016}. However, even if the developed interlocking model seems to capture well the experimental data, we would like to stress that it implicitly assumes sufficiently long chains for proper conformational statistics to be achieved, which is only approximate in our experiments. Beyond the intrinsic glassy-polymer-mimetic feature of dense granular chain assemblies, this study embodies a new illustration of friction in complex assemblies, with practical implications for new materials, textiles~\cite{bayman, keller}, and biology~\cite{ghosal,ward}.

This work was supported by the ARC Mecafood project at UMONS and the PDR project ``Capture biomim\'etique de fluides'' of the FRS-FNRS. T.S. acknowledges funding from the Global Station for Soft Matter, a project of Global Institution for Collaborative Research and Education at Hokkaido University. P.D. acknowledges the Joliot and Total chairs from ESPCI Paris.

\end{document}